\documentclass{article}
\usepackage{spconf,amsmath,graphicx}
\allowdisplaybreaks[4]
\usepackage{algorithm}
\usepackage{algorithmicx}
\usepackage{algpseudocode}
\usepackage{amsmath,amsfonts,amssymb,mathrsfs}
\usepackage{cite}
\usepackage{bm}
\usepackage[labelformat=simple]{subcaption}
\usepackage[shortlabels]{enumitem}

\everydisplay={%
	\abovedisplayskip=.3pt%
	\belowdisplayskip=.3pt%
	\abovedisplayshortskip=.3pt%
	\belowdisplayshortskip=.3pt%
}

\setlist{nolistsep,leftmargin=1em}

\def\suo{\hspace{-1pt}}
\title{Joint Transmit Beamforming and Reflection Optimization for Beyond Diagonal RIS Aided Multi-Cell MIMO Communication}
\name{Shuo Zheng, Shuowen Zhang
	}
\address{EEE Department, The Hong Kong Polytechnic University\\
E-mails: shuo.zheng@connect.polyu.hk, shuowen.zhang@polyu.edu.hk}
\begin{document}
\ninept
\maketitle
\begin{abstract}
The sixth-generation (6G) wireless  networks will rely on ultra-dense multi-cell deployment to meet the high rate and connectivity demands. However, frequency reuse leads to severe inter-cell interference, particularly for cell-edge users, which limits the communication performance. To overcome this challenge, we investigate a beyond diagonal reconfigurable intelligent surface (BD-RIS) aided multi-cell multi-user downlink MIMO communication  system, where a BD-RIS is deployed to enhance desired signals and suppress both intra-cell and inter-cell interference. We formulate the joint optimization problem of the transmit beamforming matrices at the BSs and the BD-RIS reflection matrix to maximize the weighted sum rate of all users, subject to the challenging unitary constraint of the BD-RIS reflection matrix and transmit power constraints at the BSs. To tackle this non-convex and difficult problem, we apply the weighted minimum mean squared error (WMMSE) method to transform the problem into an equivalent tractable form, and propose an efficient alternating optimization (AO) based algorithm to iteratively update the transmit beamforming and BD-RIS reflection using Lagrange duality theory and manifold optimization. Numerical results demonstrate the superiority of the proposed design over various benchmark schemes, and provide useful practical insights on the BD-RIS deployment strategy for multi-cell systems.
\end{abstract}

\section{Introduction}\vspace{-.3cm}
\label{sec:intro}

The evolution towards the sixth-generation (6G) wireless networks calls for ultra-dense multi-cell deployments to support ubiquitous connectivity and high-rate applications. In such scenarios, frequency reuse across neighboring cells is inevitable to meet the stringent communication requirements, which, however, also results in inter-cell interference. This is particularly critical for cell-edge users, which often experience weak desired signals while suffering from strong interference from neighboring cells.
To address this problem, various interference management strategies have been investigated, among which reconfigurable intelligent surface (RIS) is a promising technology due to its capability of reconfiguring the wireless propagation environment \cite{RIStutorial,RIScapacity,RISregion,myTVT}. By intelligently adjusting the phase shifts of signals impinging on its passive reflecting elements, RIS can enhance desired signal strength and suppress interference, thereby increasing the rate of multi-cell communication \cite{multicell}.

\begin{figure}
	
	\centering
	\hspace{.5cm}\includegraphics[width=.85\linewidth]{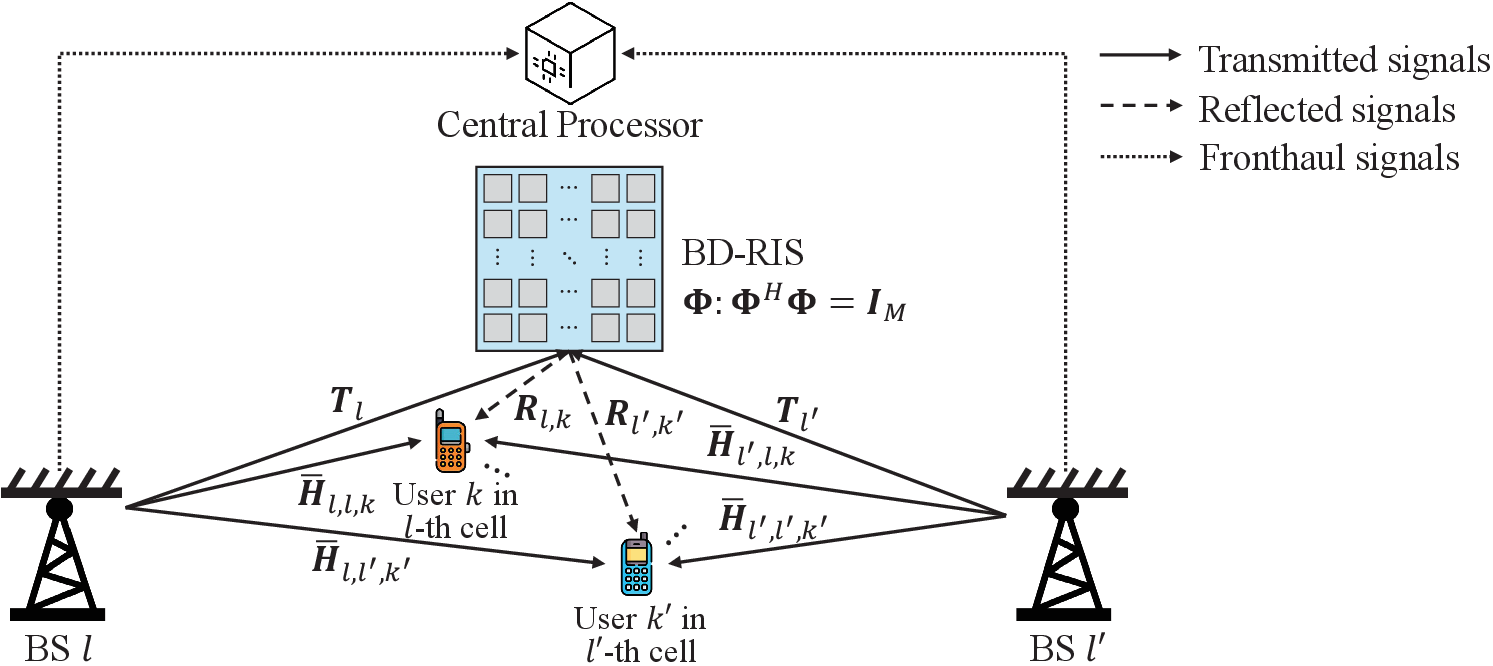}\vspace{-.25cm}
	\caption{Illustration of a BD-RIS aided multi-cell multi-user  downlink MIMO communication  system.}
	\label{model}
	\vspace{-.55cm}
\end{figure}

However, conventional (diagonal) RIS adopts a single-connected architecture that only supports diagonal reflection matrices, which restricts its beamforming flexibility and consequently limits the achievable performance gains. Recently, a novel RIS architecture, termed \textit{beyond diagonal reconfigurable intelligent surface (BD-RIS)}, has been proposed as an advancement beyond conventional RIS \cite{modeling}. Specifically, BD-RIS is capable of realizing reflection matrices beyond the diagonal structure with greater design flexibility enabled by an interconnected impedance network \cite{BDmagazine, BDtutorial}. 
 Motivated by these advantages, several works have studied the BD-RIS reflection  design in various scenarios  \cite{closedform,myTCCN,yuanye, zxq1,zxq2}. 
 While these works have demonstrated the potential of BD-RIS, most of them focused on single-cell systems. For multi-cell systems, prior efforts either considered assigning a dedicated BD-RIS to each base station (BS) \cite{multicellBDRIS1,multicellBDRIS2,multicellBDRIS4}, which leads to high deployment cost, or relied on frequency division across cells \cite{multicellBDRIS3}, which may cause resource underutilization. These limitations motivate our study in this work to unlock the full potential of BD-RIS in general multi-cell multi-user multiple-input multiple-output (MIMO) communication systems.

 In this paper, we study a BD-RIS aided multi-cell multi-user downlink MIMO communication system, where each cell consists of a multi-antenna BS and multiple multi-antenna users and different BSs are connected via fronthaul links through a central processor. One BD-RIS is deployed to enhance the multi-cell communication performance. We aim to maximize the weighted sum rate of all users via joint optimization of the transmit beamforming matrices at all BSs and the BD-RIS reflection matrix, subject to the challenging unitary constraint of the BD-RIS reflection matrix and the transmit power constraints. This problem is non-convex and difficult to solve due to the complex rate expression under both intra-cell and inter-cell interference as well as the couplings among optimization variables. By leveraging the weighted minimum mean squared error (WMMSE) method, we transform the problem into an equivalent form, based on which we further propose an alternating optimization (AO) based algorithm that iteratively updates the transmit beamforming matrices using Lagrange duality theory and the BD-RIS reflection matrix using manifold optimization. It is shown via numerical results that our proposed design is effective in multi-cell multi-user interference management and achieves superior rate performance over various benchmark schemes. It is also observed that under a given number of elements, deploying a large centralized BD-RIS at the cell edge yields higher rate over deployment of multiple smaller-sized distributed BD-RISs each near one BS.

\vspace{-.52cm}\section{System Model}\vspace{-.34cm}
\label{sec:format}

	We consider a multi-cell  multi-user downlink MIMO communication system with $L\geq 1$ cells. In each cell, one BS equipped with $N_t\geq 1$ transmit antennas aims to communicate with $K\geq 1$ users, each of which is equipped with $N_r\geq 1$ receive antennas. Different BSs are connected via fronthaul links through a central processor. An $M$-element fully-connected  BD-RIS operating in the reflective mode is deployed to enhance the multi-cell multi-user communication performance, as illustrated in Fig. \ref{model}. Let $\bm{\Phi}\in \mathbb{C}^{M\times M}$ denote the BD-RIS reflection matrix, which is subjected to the constraint $\bm{\Phi}^H\bm{\Phi}=\bm{I}_M$ due to the lossless property of the BD-RIS.
	
Let $\bm{s}_{l,k}\in \mathbb{C}^{N_s\times 1}$ denote the information symbol vector for the $k$-th user in the $l$-th cell, where $N_s$ denotes the maximum number of data streams among all users. The non-zero elements in $\bm{s}_{l,k}$ are independent of each other and follow the circularly symmetric complex Gaussian (CSCG) distribution with zero mean and unit variance. 
  Considering linear beamforming at the BS transmitters, the transmitted signal vector at each $l$-th cell,  $\bm{x}_l\in \mathbb{C}^{N_t\times 1},$ is modeled as
	$\bm{x}_l=\sum_{k=1}^K \bm{F}_{l,k}\bm{s}_{l,k}$,
where $\bm{F}_{l,k}\in\mathbb{C}^{N_t\times N_s}$ denotes the transmit beamforming matrix for the $k$-th user in the $l$-th cell. Let $P_l$ denote the transmit power budget of the $l$-th BS, which yields $\sum_{k=1}^K \|\bm{F}_{l,k}\|_\mathrm{F}^2\leq P_l$.

The signals sent from each $l$-th cell will generally arrive at users in all cells through both the direct channel and a reflected channel via the BD-RIS. Let $\bar{\bm{H}}_{l',l,k}\in\mathbb{C}^{N_r\times N_t}$, $\bm{T}_{l'}\in\mathbb{C}^{M\times N_t}$, and $\bm{R}_{l,k}\in\mathbb{C}^{N_r\times M}$ denote the direct channel from the $l'$-th BS to the $k$-th user in the $l$-th cell, the channel from the $l'$-th BS to the BD-RIS, and the channel from the BD-RIS to the $k$-th user in the $l$-th cell, respectively. We assume that these channels are perfectly known at the BSs via channel estimation techniques \cite{wangrui2, wangrui1, lhychannel} and fronthaul links connecting the BSs. The effective channel from the $l'$-th BS to the $k$-th user in the $l$-th cell is given by
\begin{align}
	\bm{H}_{l',l,k}=\bar{\bm{H}}_{l',l,k}+\bm{R}_{l,k}\bm{\Phi}\bm{T}_{l'}. 
\end{align}
Note that each user receives its desired signals while suffering from intra-cell and inter-cell interference. The received signal vector at the $k$-th user in the $l$-th cell is thus given by\vspace{-1pt}
\begin{align}
	\!\!\!\!\bm{y}_{l,k}&\!=\!\bm{H}_{l,l,k}\bm{x}_{l}+\sum\limits_{\substack{l'=1\\ l'\neq l}}^L\bm{H}_{l',l,k}\bm{x}_{l'}+\bm{n}_{l,k}\\[-4pt]
&\!=\!\bm{H}_{l,l,k}\bm{F}_{l,k}\bm{s}_{l,k} + \!\!\!\sum\limits_{(l',k')\neq (l,k)} \!\!\!\bm{H}_{l',l,k}\bm{F}_{l',k'}\bm{s}_{l',k'}\!+\!\bm{n}_{l,k},\!\!\!
\end{align}
where $\bm{n}_{l,k}\sim\mathcal{CN}(\bm{0},\sigma^2 \bm{I}_{N_r})$ denotes the CSCG noise vector at the $k$-th user in the $l$-th cell with $\sigma^2$ denoting the average noise power at each receive antenna.
The interference-plus-noise covariance matrix at the receiver of the $k$-th user in the $l$-th cell is thus expressed as
\begin{align}
	\bm{\Upsilon}_{l,k}&=\!\!\sum\limits_{(l',k')\neq (l,k)}\!\! \bm{H}_{l',l,k}\bm{F}_{l',k'}\bm{F}_{l',k'}^H\bm{H}_{l',l,k}^H+\sigma^2\bm{I}_{N_r}.
\end{align}
The achievable rate of the $k$-th user in the $l$-th cell in bits per second per Hertz (bps/Hz) is thus  given by
\begin{align}
	\!\!R_{l,k}(\bm{F},\bm{\Phi})\!=\!\log_2\det(\bm{I}_{N_r}\!+\!\bm{H}_{l,l,k}\bm{F}_{l,k}\bm{F}_{l,k}^H\bm{H}_{l,l,k}^H\bm{\Upsilon}_{l,k}^{-1}),\!\!\label{rate}
\end{align}
	where $\bm{F}\triangleq \{\bm{F}_{l,k},\ \forall (l,k)\}$ denotes the collection of transmit beamforming matrices at all BSs. 
	It is worth noting that the achievable rate for all users are dependent on the transmit beamforming matrices for all users in all cells as well as the common BD-RIS reflection matrix, which motivates our investigation on their joint optimization to maximize the multi-cell multi-user communication performance.

\section{Problem Formulation}\vspace{-.3cm}
We aim to maximize the weighted sum rate of all users by jointly optimizing the BS transmit beamforming matrices in $\bm{F}$ and the BD-RIS reflection matrix $\bm{\Phi}$, subject to the unitary constraint on the BD-RIS reflection matrix and the transmit power constraint at each BS. The optimization problem is formulated as
\begin{align}
\text{(P1)}\ \max_{\bm{F},\bm{\Phi}}\quad & \sum\limits_{l=1}^L \sum\limits_{k=1}^K \alpha_{l,k} R_{l,k}(\bm{F},\bm{\Phi})\\\
	\mathrm{s.t.}\quad  & \bm{\Phi}^H\bm{\Phi}=\bm{I}_{M}\label{uni1}\\
	& \sum\limits_{k=1}^K \Vert \bm{F}_{l,k}\Vert_\mathrm{F}^2\leq P_l,\ l=1,...,L,\label{power_cons}
\end{align}
where $\alpha_{l,k}$ denotes the weight coefficient for user $k$ in cell $l$.

It is worth noting that (P1) is a non-convex optimization problem due to the non-concave objective function over the transmit beamforming matrices $\bm{F}_{l,k}$'s and the BD-RIS reflection matrix $\bm{\Phi}$, and the non-convex unitary constraint in \eqref{uni1}. Moreover, (P1) is particularly difficult to solve since the BD-RIS reflection matrix affects both the desired signal vectors as well as the intra-cell and inter-cell interference signal vectors, and are also coupled with the transmit beamforming matrices in different cells. The complex interference in the multi-cell system and multi-antenna nature of the user receivers also make the achievable rate expressions in (\ref{rate}) challenging to deal with. 

In the following, we apply the WMMSE method to equivalently simplify the rate expressions, and develop an AO-based algorithm for finding a high-quality suboptimal solution to (P1). 

\vspace{-.43cm}\section{Proposed Solution to (P1)}\label{Sec_WMMSE}\vspace{-.28cm}
In this section, we first equivalently transform (P1) into a more tractable form via the WMMSE method. Then, we propose an AO-based algorithm for the equivalent problem where the transmit beamforming matrices and the BD-RIS reflection matrix are iteratively updated via convex optimization or manifold optimization.

\vspace{-.35cm}\subsection{WMMSE-based Equivalent Transformation of (P1)}\vspace{-.15cm}
First, let $\bm{U}_{l,k}^H\in\mathbb{C}^{N_s\times N_r}$ denote the receiver decoding matrix at the $k$-th user of the $l$-th cell. 
The decoded information symbol vector at the $k$-th user in the $l$-th cell is given by $\hat{\bm{s}}_{l,k}=\bm{U}_{l,k}^H\bm{y}_{l,k}$.
The corresponding mean squared error (MSE) matrix is thus given by
\begin{align}
	&\bm{E}_{l,k}\triangleq \mathbb{E}[(\hat{\bm{s}}_{l,k}-\bm{s}_{l,k})(\hat{\bm{s}}_{l,k}-\bm{s}_{l,k})^H]\\
	=&(\suo\bm{U}_{l,k}^H \bm{H}_{l,l,k}\bm{F}_{l,k}\!\suo-\!\suo\bm{I}_{N_s}\suo)\suo(\suo\bm{U}_{l,k}^H \bm{H}_{l,l,k}\bm{F}_{l,k} \!\suo-\!\suo\bm{I}_{N_s}\suo)^{\suo H}\!\suo+\!\suo\sigma^2\bm{U}_{l,k}^H\bm{U}_{l,k}\nonumber\\
	&\!+ \!\!\sum\limits_{(l',k')\neq (l,k)}\!\!\! \bm{U}_{l,k}^H\bm{H}_{l',l,k}\bm{F}_{l',k'}\bm{F}_{l',k'}^H\bm{H}_{l',l,k}^H\bm{U}_{l,k} .
\end{align}
Then, we introduce a set of auxiliary matrices denoted by $\bm{W}_{l,k}\succeq\bm{0},\ \forall(l,k)$.  We further define $\bm{W}\triangleq \{\bm{W}_{l,k},\ \forall (l,k)\}$ and $\bm{U}\triangleq \{\bm{U}_{l,k},\ \forall (l,k)\}$.
Based on this, (P1) can be shown to be equivalent to the following problem \cite{WMMSE}:
\begin{align}
	\!\!\suo\text{(P2)} \!\!\min_{\substack{\bm{\Phi}:\eqref{uni1},\bm{F}:\eqref{power_cons}, \bm{U}\\ \bm{W}:\bm{W}_{l,k}\succeq \bm{0},\forall l, \forall k}}\suo&\sum\limits_{l=1}^L \! \sum\limits_{k=1}^K\suo \alpha_{l,k}\suo(\mathrm{tr}(\bm{W}_{l,k}\bm{E}_{l,k})\!-\!\log_2\suo\det(\bm{W}_{l,k})\suo).\!\!\!\suo\nonumber\\[-10pt]
\end{align}

Note that (P2) is in a more tractable form than (P1) due to the absence of the interference-plus-noise covariance matrix inversion. In the following, we propose an AO-based algorithm to iteratively update the decoding matrices in $\bm{U}$, the auxiliary matrices in $\bm{W}$, as well as the transmit beamforming matrices in $\bm{F}$ and the BD-RIS reflection matrix $\bm{\Phi}$ with the other variables being fixed at each time.

\vspace{-.4cm}\subsection{Optimal Solution of $\bm{U}$ or $\bm{W}$ for (P2)}\vspace{-.21cm}

With given $\bm{F}$, $\bm{W}$, and $\bm{\Phi}$, the optimal solution of each  $\bm{U}_{l,k}$ can be obtained according to the first-order optimality as
\begin{align}
	\!\!\bm{U}_{l,k}^\star=\bm{J}_{l,k}^{-1}\!\bm{H}_{l,l,k}\bm{F}_{l,k},\ l=1,...,L, k=1,...,K,\label{Uopt}
\end{align}
where $\bm{J}_{l,k}\triangleq\bm{\Upsilon}_{l,k}+\bm{H}_{l,l,k}\bm{F}_{l,k}\bm{F}_{l,k}^{H}\bm{H}_{l,l,k}^{H}$. 

With given $\bm{F}$, $\bm{U}$, and $\bm{\Phi}$, the optimal  solution of each  $\bm{W}_{l,k}$ can be obtained by the first-order optimality condition as
\begin{align}
	\bm{W}_{l,k}^\star=\bm{E}_{l,k}^{-1},\ l=1,...,L, k=1,...,K.\label{Wopt}
\end{align}

\vspace{-.45cm}
\subsection{Optimal Solution of $\bm{F}$ for (P2)}\vspace{-.25cm}
With given $\bm{\Phi},\bm{W}$, and $\bm{U}$, (P2) reduces to the following problem:
\begin{align}
	\text{(P2-$\bm{F}$)}\ \min_{\bm{F}:\eqref{power_cons}}\quad  &\sum\limits_{l=1}^L \sum\limits_{k=1}^K \alpha_{l,k}\mathrm{tr}(\bm{W}_{l,k}\bm{E}_{l,k}).
\end{align}
Let $\bm{F}_l\triangleq \{\bm{F}_{l,k},\ \forall k\}$ denote the collection of transmit beamforming matrices at the $l$-th BS. We then have $\sum_{l=1}^L \sum_{k=1}^K \alpha_{l,k}\mathrm{tr}(\bm{W}_{l,k}\bm{E}_{l,k})$ $=\sum_{l=1}^L f_l(\bm{F}_l)+\text{constants}$, where $f_l(\bm{F}_l)$ is only dependent on the transmit beamforming matrices in $\bm{F}_l$ for the $l$-th BS and is given by\vspace{-.5pt}
\begin{align}
	f_{l}&(\bm{F}_{l})=\sum\limits_{k=1}^K\Big[\alpha_{l,k}\mathrm{tr}\big(\bm{W}_{l,k}(\bm{U}_{l,k}^H \bm{H}_{l,l,k}\bm{F}_{l,k}\!-\!\bm{I}_{N_s})\nonumber\\[-2.8pt]
	&\!\!\times (\bm{U}_{l,k}^H \bm{H}_{l,l,k}\bm{F}_{l,k} \!-\!\bm{I}_{N_s})^H\big)+\!\!\!\!\!\sum\limits_{(l',k')\neq(l,k)}\!\!\! \alpha_{l',k'}\mathrm{tr}\big(\bm{W}_{l',k'}\bm{U}_{l',k'}^H\nonumber\\[-2.8pt]
	& \hspace{2cm}\times\bm{H}_{l,l',k'}\bm{F}_{l,k}\bm{F}_{l,k}^H\bm{H}_{l,l',k'}^H\bm{U}_{l',k'}\big)\Big].
\end{align}
By further noting that $\bm{F}_l$'s are not coupled in the constraints of (P2-$\bm{F}$), (P2-$\bm{F}$) can be equivalently solved by solving the following problem for $l=1,...,L$:\vspace{-.3pt}
\begin{align}
	\text{(P2-$\bm{F}_l$)}\quad \min_{\bm{F}_l:\sum_{k=1}^K \Vert \bm{F}_{l,k}\Vert_\mathrm{F}^2\leq P_l}\ &f_{l}(\bm{F}_{l}).
\end{align}
Problem (P2-$\bm{F}_l$) is convex and can be solved via the interior-point method or existing software such as CVX.

It is worth noting that the optimal  solution can also be derived in semi-closed form as follows via the Lagrange duality theory.
Let $\mu_l\geq 0$ denote the dual variable associated with the transmit power constraint of (P2-$\bm{F}_l$). The Lagrangian of (P2-$\bm{F}_l$) is given by
$\mathcal{L}(\bm{F}_l,\suo\mu_l)\!=f_{l}(\bm{F}_{l})+\mu_l (\sum_{k=1}^K \Vert \bm{F}_{l,k}\Vert_\mathrm{F}^2-P_l)$.
Based on the first-order optimality condition, 
the optimal solution to each  $\bm{F}_{l,k}$ can be derived in semi-closed form as\vspace{-.3pt}
\begin{align}
	\!\!\!\bm{F}_{l,k}^\star=\,&\alpha_{l,k}\Big(\sum\limits_{l'=1}^L\!\sum\limits_{k'=1}^K \alpha_{l',k'} \bm{H}_{l,l',k'}^H \bm{U}_{l',k'}\bm{W}_{l',k'}\bm{U}_{l',k'}^H \bm{H}_{l,l',k'}\nonumber\\[-1.2pt]
	& \! +\!\mu_l^\star\bm{I}_{N_t}\Big)^{\!-1} \bm{H}_{l,l,k}^H \bm{U}_{l,k}\bm{W}_{l,k},\  l\!=\!1,...,L, k\!=\!1,...,K,\label{Fopt}\!\!
\end{align}
where each $\mu_l^\star$ satisfies the complement slackness condition $\mu_l^\star$ $\times(\sum_{k=1}^K \Vert \bm{F}_{l,k}^\star\Vert_\mathrm{F}^2-P_l)=0$. Specifically, if $\sum_{l'=1}^L\sum_{k'=1}^K \alpha_{l',k'}$ $\times \bm{H}_{l,l',k'}^H \bm{U}_{l',k'}\bm{W}_{l',k'}\bm{U}_{l',k'}^H \bm{H}_{l,l',k'}$ is invertible and $\bm{F}_{l,k}^\star$ in (\ref{Fopt}) with $\mu_l^\star=0$ satisfies the transmit power constraint, we have $\mu_l^\star=0$. Otherwise, we need to find $\mu_l^\star>0$ such that $\sum_{k=1}^K \|\bm{F}_{l,k}(\mu_l^\star)\|_\mathrm{F}^2= P_l$. Denote $\bm{Q}_{l}\triangleq\sum_{l'=1}^L\sum_{k'=1}^K \alpha_{l',k'}\bm{H}_{l,l',k'}^H \bm{U}_{l',k'}\bm{W}_{l',k'}\bm{U}_{l',k'}^H $ $\times\bm{H}_{l,l',k'}$ and $\bm{D}_l\bm{\Lambda}_l\bm{D}_l^H=\bm{Q}_{l}$ as its eigenvalue decomposition (EVD). 
 We further denote $\bm{C}_l=\sum_{k=1}^K \alpha_{l,k}^2\bm{D}_l^H\bm{H}_{l,l,k}^H \bm{U}_{l,k}\bm{W}_{l,k}$ $\times\bm{W}_{l,k}^H\bm{U}_{l,k}^H \bm{H}_{l,l,k}\bm{D}_l$. Then, the complement slackness condition can be equivalently written as \vspace{-1.9pt}
\begin{align}
\mathrm{tr}((\bm{\Lambda}_l+\mu_l^\star \bm{I}_{N_t})^{-2} \bm{C}_l)=\sum\limits_{n_t=1}^{N_t} \frac{{\bm{C}_l}_{n_t,n_t}}{({\bm{\Lambda}_l}_{n_t,n_t}+\mu_l^\star)^2}= P_l.\\[-17pt]\nonumber
\end{align}
Since $\sum_{n_t=1}^{N_t} \frac{{\bm{C}_l}_{n_t,n_t}}{({\bm{\Lambda}_l}_{n_t,n_t}+\mu_l^\star)^2}$ is a decreasing function over $\mu_l^\star>0$,
 $\mu_l^\star$ can be efficiently obtained by using the bi-section method.

\vspace{-.45cm}\subsection{Optimization of $\bm{\Phi}$ for (P2)}\label{optimizePhi}\vspace{-.25cm}
With given $\bm{F}$, $\bm{W}$, and $\bm{U}$, (P2) is reduced to the following problem:
\begin{align}
	\text{(P2-$\bm{\Phi}$)}\  \min_{\bm{\Phi}:\eqref{uni1}}\quad &\sum\limits_{l=1}^L \sum\limits_{k=1}^K \alpha_{l,k}\mathrm{tr}(\bm{W}_{l,k}\bm{E}_{l,k}).
\end{align}
By extracting the terms related to $\bm{\Phi}$,
we can re-express the objective in a more compact form as
$\sum_{l=1}^L\sum_{k=1}^K \alpha_{l,k}\mathrm{tr}(\bm{W}_{l,k}\bm{E}_{l,k}) =\mathrm{tr}(\bm{A}_1\bm{\Phi}\bm{A}_2\bm{\Phi}^H)\!+\!2\mathfrak{Re}\{\mathrm{tr}((\bm{B}_1\!+\!\bm{B}_2)\bm{\Phi})\}\!+\!\text{constants}$, 
where
\begin{align}
	\bm{A}_1&=\sum\limits_{l=1}^L\sum\limits_{k=1}^K \alpha_{l,k}\bm{R}_{l,k}^H\bm{U}_{l,k}\bm{W}_{l,k}\bm{U}_{l,k}^H\bm{R}_{l,k}\label{A1}\\[-2pt]
	\bm{A}_2&=\sum\limits_{l'=1}^L\sum\limits_{k'=1}^K \bm{T}_{l'}\bm{F}_{l',k'}\bm{F}_{l',k'}^H\bm{T}_{l'}^H\\
	\bm{B}_1&=\sum\limits_{l=1}^L\!\sum\limits_{k=1}^K\!\sum\limits_{l'=1}^L\!\sum\limits_{k'=1}^K \!\alpha_{l,k}\bm{T}_{l'}\bm{F}_{l',k'}\bm{F}_{l',k'}^H \bar{\bm{H}}_{l',l,k}^H\nonumber\\
	&\hspace{3.8cm}\times \bm{U}_{l,k}\bm{W}_{l,k}\bm{U}_{l,k}^H\bm{R}_{l,k}\\
	\bm{B}_2&=-\sum\limits_{l=1}^L\sum\limits_{k=1}^K\alpha_{l,k}\bm{T}_l\bm{F}_{l,k}\bm{W}_{l,k}\bm{U}_{l,k}^H\bm{R}_{l,k}.\label{B2}
\end{align}
Then, (P2-$\bm{\Phi}$) can be equivalently rewritten as
\begin{align}
	\!\!\!\!\text{(P2'-$\bm{\Phi}$)}\ \  \min_{\bm{\Phi}:\eqref{uni1}}\ \  &\mathrm{tr}(\bm{A}_1\bm{\Phi}\bm{A}_2\bm{\Phi}^H\suo)\!\suo+\!\suo2\mathfrak{Re}\{\mathrm{tr}((\bm{B}_1\!\suo+\suo\!\bm{B}_2)\bm{\Phi})\}.\!\!\!\!
\end{align}
Note that (P2'-$\bm{\Phi}$) is still non-convex due to the unitary constraint in \eqref{uni1}. 
However, notice that $\bm{\Phi}$ belongs to an $M\times M$ complex Stiefel manifold represented by
$\mathrm{St}_{M} =\{\bm{\Phi}\in\mathbb{C}^{M\times M}: \bm{\Phi}^H\bm{\Phi}=\bm{I}_{M}\}$.
Therefore, to tackle (P2'-$\bm{\Phi}$), we can employ a manifold-based algorithm to iteratively update $\bm{\Phi}$ \cite{steepest}. Specifically, we first calculate the Euclidean gradient as
\begin{align}
	\bm{\nabla}&=\frac{\partial g(\bm{\Phi})}{\partial \bm{\Phi}^*}=\bm{A}_{1}\bm{\Phi}\bm{A}_{2}+\bm{B}_{1}^H+\bm{B}_{2}^H.\label{Egradient}
\end{align}
Based on this, the Riemannian gradient can be obtained as
\begin{align}
	\widetilde{\bm{\nabla}}=\bm{\nabla}-\bm{\Phi}{\bm{\nabla}}^H\bm{\Phi}.\label{Rgradient}
\end{align}
Then, we translate the Riemannian gradient to identity for further calculating the rotation matrix:
\begin{align}
	\widehat{\bm{\nabla}}=\widetilde{\bm{\nabla}}{\bm{\Phi}}^H=\bm{\nabla}{\bm{\Phi}}^H-\bm{\Phi}{\bm{\nabla}}^H.\label{identity}
\end{align}
Based on this, the rotation matrix can be obtained as follows, which provides the
geodesic motion along the steepest descent direction:
\begin{align}
	\bm{P}=\exp(-\xi \widehat{\bm{\nabla}}),\label{rotation}
\end{align}
with $\xi$ denoting the step length that can be tuned based on the Armijo rule.
Finally, $\bm{\Phi}$ can be updated as follows:
\begin{align}
	\bm{\Phi}=\bm{P}\bm{\Phi}.\label{updatePhig}
\end{align}

By repeating these procedures and iteratively optimizing $\bm{\Phi}$, we can obtain a high-quality suboptimal solution to Problem (P2'-$\bm{\Phi}$). 
 According to \cite{steepest}, the manifold-based algorithm for (P2'-$\bm{\Phi}$) is guaranteed to converge. The computational complexity of the manifold-based algorithm is given by $\mathcal{O}(I_{\mathrm{MF}}I_{\xi}M^3)$, where $I_{\mathrm{MF}}$ and $I_{\xi}$ denote the numbers of iterations for the geodesic steepest descent and step size adjustment, respectively.
	\begin{algorithm}[t]
		\caption{Proposed Algorithm for (P1)}
		\label{algWMMSE}
		\begin{algorithmic}[1]
			\Repeat
			\State Update $\bm{U}$, $\bm{W}$, and $\bm{F}$ according to \eqref{Uopt}, \eqref{Wopt}, and \eqref{Fopt}, respectively.
			\Repeat
			\State Calculate $\bm{\nabla}$, $\widetilde{\bm{\nabla}}$, $\widehat{\bm{\nabla}}$, and $\bm{P}$ according to \eqref{Egradient}, \eqref{Rgradient}, \eqref{identity}, and \eqref{rotation}, respectively.
			\State Update $\bm{\Phi}$ according to \eqref{updatePhig}.
			\Until{convergence.}
			\Until{convergence.}
		\end{algorithmic}
	\end{algorithm}

\vspace{-.4cm}\subsection{Overall Algorithm for (P1)}\vspace{-.15cm}
	To summarize, the overall algorithm for Problem (P1) via solving (P2) is summarized in Algorithm \ref{algWMMSE}. Since the objective function of (P2) is lower-bounded and each update of optimization variables leads to a non-increasing objective value, the overall algorithm is guaranteed to converge. By noting that $N_s\leq \min\{N_t,N_r\}$, 
	 the worst-case computational complexity for Algorithm 1 can be shown to be
$\mathcal{O}(I_{\mathrm{AO}}(L^2K^2(N_t N_r^2+N_t N_r M + N_r M^2)+LK(N_r^3 + N_t^2 N_r +  N_r^2 M +  N_t^2 M) + L N_t^3 + I_\mathrm{MF}I_{\xi}M^3))$,
	  where $I_{\mathrm{AO}}$ denotes the total number of AO iterations.

\vspace{-.52cm}\section{Numerical Results}\vspace{-.38cm}
In this section, we provide numerical results to evaluate the performance of the proposed joint transmit beamforming and BD-RIS reflection design. We set $L=2$, $K=2$, $N_t=4$, $N_r=2$, $N_s=2$, $M=20$, $P_l=P=30$ dBm,$\ l=1,...,L$, and $\sigma^2=-104$ dBm unless otherwise specified \cite{multicell}. The coordinates of the two BSs and the BD-RIS are set as $[0,0]$ m, $[600,0]$ m, and $[300,0]$ m, respectively. Two users in each cell are randomly and  uniformly distributed in a disk of radius $20$ m centered at $[280,0]$ m; or $[320,0]$ m, respectively. The direct BS-user channels are modeled under Rayleigh fading with path loss exponent $3.75$. The  channels related to the BD-RIS are modeled under  Rician fading with path loss exponent $2.2$ and Rician factor $3$. The path loss at reference distance $1$ m is set as $30$ dB. We set the weight coefficient for all users as $\alpha_{l,k}=1$.

\begin{figure}[t]
	\centering
	\includegraphics[width=.7\linewidth]{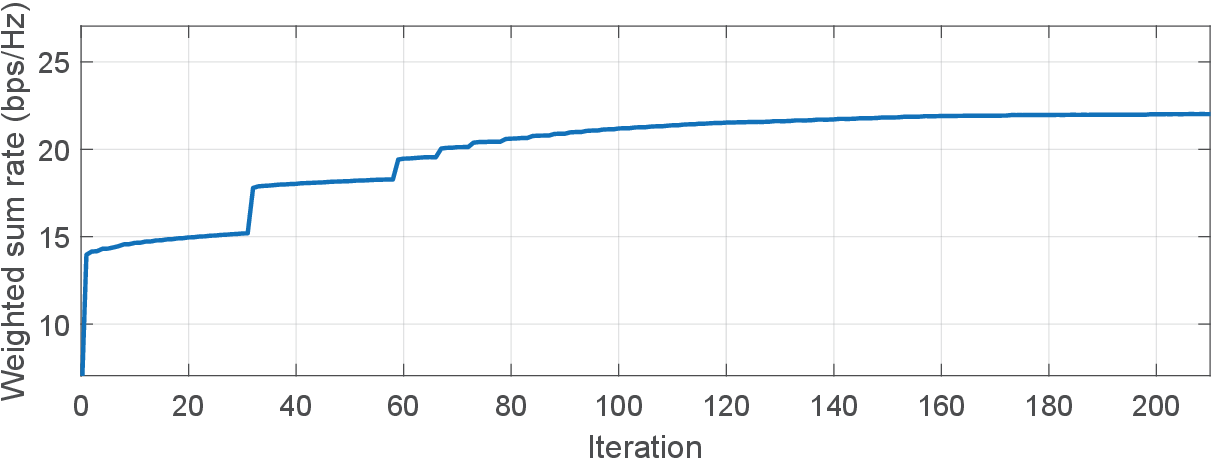}\vspace{-.25cm}
	\caption{Convergence behavior of Algorithm \ref{algWMMSE}.}
	\label{outer_convergence}
	\vspace{-.3cm}
\end{figure}

\begin{figure}[t]
	\centering
	\begin{minipage}[b]{.495\linewidth}
		\centering
		\includegraphics[width=\linewidth]{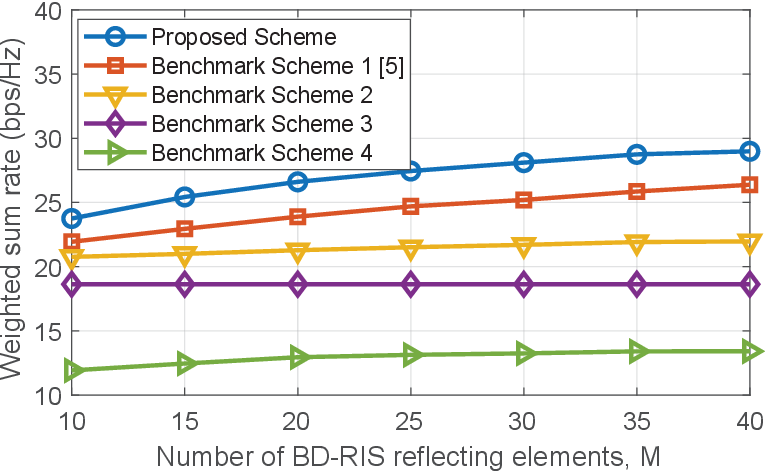}
		\subcaption{Weighted sum rate vs. $M$.}
		\label{rate_vsM}
	\end{minipage}
	\hfill
	\begin{minipage}[b]{0.495\linewidth}
		\centering
		\includegraphics[width=\linewidth]{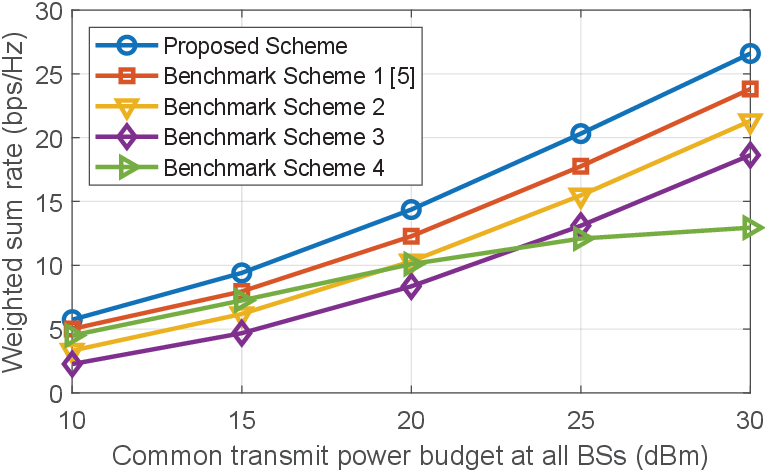}
		\subcaption{Weighted sum rate vs. power.}
		\label{rate_vsP}
	\end{minipage}
	\vspace{-.45cm}
	\caption{Weighted sum rate versus number of BD-RIS elements or BS transmit power.}
	\label{fig:vs}
	\vspace{-.35cm}
\end{figure}
\begin{figure}[t]
	\centering
	\includegraphics[width=0.57\linewidth]{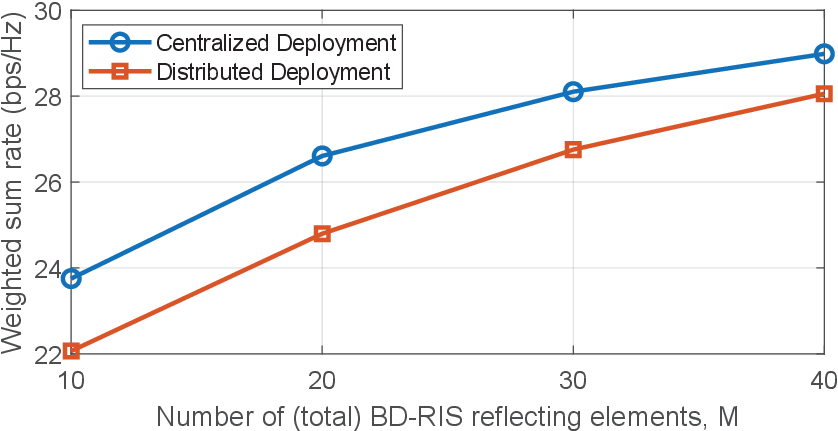}
	\vspace{-.15cm}
	\caption{Weighted sum rate with different BD-RIS deployment.}
	\label{dis}
	\vspace{-.3cm}
\end{figure}

Firstly, we show the convergence behavior of Algorithm 1 in Fig. \ref{outer_convergence}. It is observed that the proposed algorithm converges fastly within 200 iterations, and the converged weighted sum rate is significantly higher than that at the initial point. This is due to the optimal solutions of $\bm{U}$, $\bm{W}$, and $\bm{F}$ or high-quality suboptimal solution of $\bm{\Phi}$ obtained in each iteration.

Then, we consider the following benchmark schemes for performance comparison:
\begin{itemize}
	\itemsep .5pt
	\item \textbf{Benchmark scheme 1}: We consider a {\bf{conventional diagonal RIS}} aided multi-cell multi-user downlink system, where $\bm{\Phi}$ is a diagonal matrix and optimized to maximize the weighted sum rate using the majorization-minimization based algorithm in \cite{multicell}.
	\item \textbf{Benchmark scheme 2}: We consider {\bf{random BD-RIS reflection}}. Specifically, we randomly generate 100 BD-RIS reflection matrices and select the one with the best performance. In each realization, we randomly generate a complex matrix $\bm{S}\in \mathbb{C}^{M\times M}$ where the real and imaginary parts of each entry is randomly generated under uniform distribution in $[0,1]$; then, we perform QR decomposition of $\bm{S}$ and obtain the Q matrix as the BD-RIS reflection matrix, which is guaranteed to be unitary.
	\item \textbf{Benchmark scheme 3}: We consider a multi-cell multi-user downlink system {\bf{without BD-RIS}}, where the transmit beamforming matrices are optimized using the same approach as in Section \ref{Sec_WMMSE}.
	\item \textbf{Benchmark scheme 4}: We consider a {\bf{non-cooperative}} multi-cell multi-user downlink system, where each BS independently designs the its transmit beamforming to maximize the weighted sum rate of users in its cell. The BD-RIS aids the communication within one cell at a time in a round-robin manner to maximize the weighted sum rate of users in the corresponding cell.
\end{itemize}
 
In Fig. \ref{rate_vsM}, we show the weighted sum rate versus the number of BD-RIS reflecting elements. It is observed that our proposed design outperforms all benchmark schemes, with an average 2.5947 bps/Hz rate gain compared with its conventional diagonal RIS aided counterpart. Moreover, as the number of BD-RIS reflecting elements increases, the performance gain of the proposed design over systems without BD-RIS, random BD-RIS reflection, and non-cooperative design increases, which demonstrates the effectiveness of the proposed BD-RIS reflection design together with the transmit beamforming design. In Fig. \ref{rate_vsP}, we show the weighted sum rate versus the common BS transmit power budget. It is observed that our proposed design outperforms all benchmark schemes. Particularly, the non-cooperative design encounters a rate floor as the transmit power increases, which is due to its lack of inter-cell interference management ability. In contrast, our proposed design yields satisfactory interference management capability due to the judicious joint optimization of the BD-RIS reflection and transmit beamforming.

Finally, note that throughout this paper, we consider a multi-cell system aided by a single BD-RIS under a \emph{centralized} deployment strategy. In Fig. \ref{dis}, we consider another \emph{distributed} deployment strategy of the BD-RIS and evaluate its performance in comparison with the centralized one. Specifically, we propose to deploy two distributed BD-RISs each with $M/2$ elements and located at $[5,0]$ m or $[595,0]$ m near one BS. The BS transmit beamforming and BD-RIS reflection under the distributed deployment are designed in a similar manner as that proposed in the paper. Note that compared to the centralized deployment where the BD-RIS is located at the cell edge, the distributed deployment strategy can enable smaller path loss for the channels between each BS and the users in its corresponding cell, due to the short distance between each BS and its adjacent BD-RIS. However, the overall reflection gain and interference management capability at each smaller-sized distributed BD-RIS is lower than that at the centralized BD-RIS. Thus, it is unclear which strategy yields better rate performance. In Fig. \ref{dis}, it is observed that the centralized deployment strategy outperforms the distributed one under the considered setup, while the performance gain decreases as the total number of reflecting elements increases. This indicates that as $M$ increases, the signal processing capability of each distributed BD-RIS enhances, such that its gap to the large centralized BD-RIS decreases under the smaller path loss.

\vspace{-.22cm}\section{Conclusions}\vspace{-.12cm}
This paper aimed to unlock the full potential of BD-RIS in a multi-cell multi-user downlink MIMO system via joint optimization of the transmit beamforming matrices at the BSs and the BD-RIS reflection matrix towards weighted sum rate maximization among all users. Despite the non-convexity of the problem and complex rate expressions in log-determinant forms, we proposed an AO-based algorithm which iteratively updates the transmit beamforming matrices via Lagrange duality theory and the BD-RIS reflection matrix via manifold optimization, thanks to the equivalent problem transformation via the WMMSE method. Numerical results showed that the proposed design outperforms various benchmark schemes in terms of weighted sum rate due to the judicious beamforming and reflection co-design. Moreover, it was unveiled that deploying one centralized BD-RIS at the cell edge achieves better performance than deploying multiple smaller-sized distributed BD-RISs each near one BS.

\newpage
\vfill\pagebreak

\bibliographystyle{IEEEbib}
\bibliography{strings,refs}

\begin{thebibliography}{10}

\bibitem{RIStutorial}
Q.~Wu, S.~Zhang, B.~Zheng, C.~You, and R.~Zhang,
\newblock ``Intelligent reflecting surface-aided wireless communications: A
  tutorial,''
\newblock {\em {IEEE} Trans. Commun.}, vol. 69, no. 5, pp. 3313--3351, May
  2021.

\bibitem{RIScapacity}
S.~Zhang and R.~Zhang,
\newblock ``Capacity characterization for intelligent reflecting surface aided
  {MIMO} communication,''
\newblock {\em {IEEE} J. Sel. Areas Commun.}, vol. 38, no. 8, pp. 1823--1838,
  Aug. 2020.

\bibitem{RISregion}
S.~Zhang and R.~Zhang,
\newblock ``Intelligent reflecting surface aided multi-user communication:
  Capacity region and deployment strategy,''
\newblock {\em {IEEE} Trans. Commun.}, vol. 69, no. 9, pp. 5790--5806, Sep.
  2021.

\bibitem{myTVT}
S.~Zheng, B.~Lv, T.~Zhang, Y.~Xu, G.~Chen, R.~Wang, and P.~C. Ching,
\newblock ``On {DoF} of active {RIS}-assisted {MIMO} interference channel with
  arbitrary antenna configurations: When will {RIS} help?,''
\newblock {\em {IEEE} Trans. Veh. Technol.}, vol. 72, no. 12, pp. 16828--16833,
  Dec. 2023.

\bibitem{multicell}
C.~Pan, H.~Ren, K.~Wang, W.~Xu, M.~Elkashlan, A.~Nallanathan, and L.~Hanzo,
\newblock ``Multicell {MIMO} communications relying on intelligent reflecting
  surfaces,''
\newblock {\em {IEEE} Trans. Wireless Commun.}, vol. 19, no. 8, pp. 5218--5233,
  Aug. 2020.

\bibitem{modeling}
S.~Shen, B.~Clerckx, and R.~Murch,
\newblock ``Modeling and architecture design of reconfigurable intelligent
  surfaces using scattering parameter network analysis,''
\newblock {\em {IEEE} Trans. Wireless Commun.}, vol. 21, no. 2, pp. 1229--1243,
  Feb. 2022.

\bibitem{BDmagazine}
H.~Li, S.~Shen, M.~Nerini, and B.~Clerckx,
\newblock ``Reconfigurable intelligent surfaces 2.0: Beyond diagonal phase
  shift matrices,''
\newblock {\em {IEEE} Commun. Mag.}, vol. 62, no. 3, pp. 102--108, 2024.

\bibitem{BDtutorial}
H.~Li, M.~Nerini, S.~Shen, and B.~Clerckx,
\newblock ``A tutorial on beyond-diagonal reconfigurable intelligent surfaces:
  Modeling, architectures, system design and optimization, and applications,''
\newblock {\em {IEEE} Commun. Surveys Tuts.}, vol. 28, pp. 4086--4126, 2026.

\bibitem{closedform}
M.~Nerini, S.~Shen, and B.~Clerckx,
\newblock ``Closed-form global optimization of beyond diagonal reconfigurable
  intelligent surfaces,''
\newblock {\em {IEEE} Trans. Wireless Commun.}, vol. 23, no. 2, pp. 1037--1051,
  Feb. 2024.

\bibitem{myTCCN}
S.~Zheng and S.~Zhang,
\newblock ``Beyond diagonal intelligent reflecting surface aided integrated
  sensing and communication,''
\newblock {\em {IEEE} Trans. on Cogn. Commun. Netw.}, vol. 11, no. 5, pp.
  2864--2878, Oct. 2025.

\bibitem{yuanye}
Y.~Yuan and S.~Zhang,
\newblock ``Beyond diagonal {IRS} aided {OFDM}: Rate maximization under
  frequency-dependent reflection,''
\newblock in {\em Proc. IEEE Global Commun. Conf. (Globecom)}, Dec. 2025.

\bibitem{zxq1}
X.~Zhang, L.~Liu, S.~Zhang, W.~Zhu, and H.~Zhang,
\newblock ``Optimizing rate-{CRB} performance for beyond diagonal
  reconfigurable intelligent surface enabled {ISAC},''
\newblock {\em {IEEE} Commun. Lett.}, vol. 29, no. 11, pp. 2506--2510, Nov.
  2025.

\bibitem{zxq2}
X.~Zhang, L.~Liu, S.~Zhang, and H.~Zhang,
\newblock ``Beyond diagonal reconfigurable intelligent surface enabled sensing:
  {Cramér-Rao} bound optimization,''
\newblock {\em {IEEE} Wireless Commun. Lett.}, vol. 14, no. 12, pp. 3852--3856,
  Dec. 2025.

\bibitem{multicellBDRIS1}
K.~D. Katsanos, P.~D. Lorenzo, and G.~C. Alexandropoulos,
\newblock ``The interference broadcast channel with reconfigurable intelligent
  surfaces: A cooperative sum-rate maximization approach,''
\newblock in {\em Proc. IEEE Int. Wkshps. Signal Process. Adv. Wireless Commun.
  (SPAWC)}, Sep. 2024, pp. 551--555.

\bibitem{multicellBDRIS2}
K.~D. Katsanos, P.~D. Lorenzo, and G.~C. Alexandropoulos,
\newblock ``Multi-{RIS}-empowered multiple access: A distributed sum-rate
  maximization approach,''
\newblock {\em {IEEE} J. Sel. Topics Signal Process.}, vol. 18, no. 7, pp.
  1324--1338, Oct. 2024.

\bibitem{multicellBDRIS4}
C.~Zhang, W.~U. Khan, A.~K. Bashir, A.~K. Dutta, A.~U. Rehman, and M.~M.~A.
  Dabel,
\newblock ``Sum rate maximization for {6G} beyond diagonal {RIS}-assisted
  multi-cell transportation systems,''
\newblock {\em {IEEE} Trans. Intell. Transp. Syst.}, vol. 26, no. 10, pp.
  17601--17611, Oct. 2025.

\bibitem{multicellBDRIS3}
A.~S. de~Sena, M.~Rasti, N.~H. Mahmood, and M.~Latva-aho,
\newblock ``Beyond diagonal {RIS} for multi-band multi-cell {MIMO} networks: A
  practical frequency-dependent model and performance analysis,''
\newblock {\em {IEEE} Trans. Wireless Commun.}, vol. 24, no. 1, pp. 749--766,
  Jan. 2025.

\bibitem{wangrui2}
R.~Wang, S.~Zhang, B.~Clerckx, and L.~Liu,
\newblock ``Low-overhead channel estimation framework for beyond diagonal
  reconfigurable intelligent surface assisted multi-user {MIMO}
  communication,''
\newblock {\em {IEEE} Trans. Signal Process.}, vol. 73, pp. 4700--4717, Nov.
  2025.

\bibitem{wangrui1}
R.~Wang, S.~Zhang, and L.~Liu,
\newblock ``Low-overhead channel estimation for beyond diagonal reconfigurable
  intelligent surface aided single-user communication,''
\newblock in {\em Proc. {IEEE} Int. Conf. Wireless Commun. Signal Process.
  (WCSP)}, Oct. 2024, pp. 305--310.

\bibitem{lhychannel}
H.~Li, S.~Shen, Y.~Zhang, and B.~Clerckx,
\newblock ``Channel estimation and beamforming for beyond diagonal
  reconfigurable intelligent surfaces,''
\newblock {\em {IEEE} Trans. Signal Process.}, vol. 72, pp. 3318--3332, Jul.
  2024.

\bibitem{WMMSE}
Q.~Shi, M.~Razaviyayn, Z.-Q. Luo, and C.~He,
\newblock ``An iteratively weighted {MMSE} approach to distributed sum-utility
  maximization for a {MIMO} interfering broadcast channel,''
\newblock {\em {IEEE} Trans. Signal Process.}, vol. 59, no. 9, pp. 4331--4340,
  Sep. 2011.

\bibitem{steepest}
T.~E. Abrudan, J.~Eriksson, and V.~Koivunen,
\newblock ``Steepest descent algorithms for optimization under unitary matrix
  constraint,''
\newblock {\em {IEEE} Trans. Signal Process.}, vol. 56, no. 3, pp. 1134--1147,
  Mar. 2008.

\end{thebibliography}

\end{document}